\documentclass[10pt,sigconf]{acmart} %

\usepackage[english]{babel}
\usepackage{blindtext}
\usepackage{graphicx}
\usepackage{xcolor,colortbl}
\usepackage{multicol}
\usepackage{multirow}
\usepackage{tcolorbox}
\usepackage{listings}
\usepackage[colorinlistoftodos]{todonotes}
\usepackage[toc,page]{appendix}
\usepackage{pgffor}
\usepackage{pgfplots}
\usepackage{multicol}
\usepackage{pgfgantt}

\definecolor{blue}{RGB}{44,123,182}
\definecolor{green}{RGB}{255,255,191}
\definecolor{orange}{RGB}{253,174,97}
\definecolor{red}{RGB}{215,25,28}

\definecolor{pblue}{rgb}{0.13,0.13,1}
\definecolor{pgreen}{rgb}{0,0.5,0}
\definecolor{pred}{rgb}{0.9,0,0}
\definecolor{pgrey}{rgb}{0.46,0.45,0.48}
\definecolor{mediumslateblue}{rgb}{0.48, 0.41, 0.93}
\definecolor{electricviolet}{rgb}{0.56, 0.0, 1.0}
\definecolor{LightCyan}{rgb}{0.88,1,1}
\definecolor{LightBlue}{rgb}{0,0,0.8}

\definecolor{DarkBlue}{rgb}{0,0,0.2}

\lstdefinestyle{myCustomIvyStyle}{
  numbers=left,
  frame=single,
  stepnumber=1,
  numbersep=10pt,
  tabsize=2,
  basicstyle=\footnotesize,
  showspaces=false,
  showstringspaces=false,
  keywordstyle=\color{pblue},
  keywordstyle=[2]\color{mediumslateblue}, 
  keywordstyle=[3]\color{electricviolet}, 
  identifierstyle=\color{black},
  commentstyle=\itshape\color{pgreen},
  stringstyle=\color{pred},
  morekeywords={
    then, after, type, String, action, object, variant, of, import, 
    instance, around, if, while, implement, else, before, ensure, 
    module, returns, return, python, implementation
  },
  morekeywords=[2]{
    require, var, call, destructor, instantiate, struct, execute, 
    is_set, value, set, ip, ipv6, interpret, virtual
  },
  morekeywords=[3]{
    this, stream_pos, stream_data, trans_params_struct, cid, 
    quic_packet_type, bool, frame, stream_id, quic_packet, endpoint, 
    microsecs, pkt_num
  },
  extendedchars=true,
  morecomment=[l]{\#} 
}

\usepackage{epigraph}
\renewcommand\footnotetextcopyrightpermission[1]{} 
\setcopyright{none}

\acmDOI{}

\acmISBN{}

\acmConference[EPIQ'21]{Virtual Conference}
\acmYear{2021}

\acmPrice{}

\begin{document}
\title{Verifying QUIC implementations using Ivy}



\author{Christophe Crochet, Tom Rousseaux, J-F Sambon, Maxime Piraux, Axel Legay}

\renewcommand{\shortauthors}{Crochet .et al.}
    
\begin{abstract}
   \texttt{QUIC} is a new transport protocol combining the reliability and congestion control features of \texttt{TCP} with the security features of \texttt{TLS}.  One of the main challenges with \texttt{QUIC} is to guarantee that any of its implementation follows the \texttt{IETF} specification. This challenge is particularly appealing as the specification is written in textual language, and hence may contain ambiguities. In a recent work, McMillan and Zuck proposed a formal representation of part of \texttt{draft-18} of the IETF specification. They also showed that this representation made it possible to efficiently generate tests to stress four implementations of \texttt{QUIC}. Our first contribution is to complete and extend the formal representation from \texttt{draft-18} to \texttt{draft-29}. Our second contribution is to test seven implementations of both \texttt{QUIC} client and server. Our last contribution is to show that our tool can highlight ambiguities in the \texttt{QUIC} specification, for which we suggest paths to corrections. 

\end{abstract}


\maketitle  

\section{Introduction}



\texttt{QUIC} is a new network protocol that is intended to make the Internet faster, more secure and more flexible. It is designed to replace the entire \texttt{TCP}/\texttt{TLS}/\texttt{HTTP} stack and is built above \texttt{UDP}. \texttt{QUIC} is now widely adopted and many other applications are made compatible with the protocol. This includes, for example, \texttt{MQTT}\cite{kumar2019implementation} or \texttt{DNS}\cite{huitema-quic-dnsoquic-07}. As \texttt{QUIC} is getting deployed, ensuring that \texttt{QUIC} implementations meet the set of requirements defined by the \texttt{QUIC} specification is critical. This specification is an English text document describing the protocol. It is discussed by the Internet Engineering Task Force (IETF) and is split across several Request For Comments (RFC) documents. The main document is \texttt{RFC9000}, containing more than 250 \texttt{MUST} statements indicating properties that must be met by all implementations~\cite{rfc9000}. As RFCs lack a formal definition, they can lead to ambiguities of understanding~\cite{10.1145/3472305.3472314}.

Several approaches have been proposed to verify that \texttt{QUIC} implementations follow the specification requirements. The most common approach, called \textit{interoperability testing}, manually generates sets of tests from the requirements and then compares \texttt{QUIC} implementations with respect to those sets. This approach has been used in the \textit{QUIC-Tracker} test suite~\cite{Piraux_2018} and the \textit{QUIC Interop Runner}~\cite{quic-interop}. Albeit such approach sounds appealing, it is limited by the capacity to manually produce interesting test suites from the requirements. Another approach \cite{10.1145/3452296.3472938} is to produce a mathematical model for the protocol and its requirements, and then use formal verification to automatically assess  correctness. This approach is sound and precise. However, as it is limited to models only, it does not guarantee that subsequent implementations fulfil the requirements.


In a series of recent work \cite{10.1145/3341302.3342087, mcmillan2019compositional}, the authors have proposed a trade off between the two approaches. They implemented a formal language to specify protocols requirements called Ivy. Any specification written in Ivy automatically generates various and well-distributed test cases that can be exercised on \texttt{QUIC} implementation. In their work, the authors verified several requirements of \texttt{draft-18} of the \texttt{QUIC} specification. However, many requirements were left unimplemented, e.g. the management of the transport errors by \texttt{QUIC} implementations.  

Our three contributions are summarized as follows.

$(1).$ We contribute to the formal specification of \texttt{QUIC} in Ivy by implementing a significant part of \texttt{draft-29} requirements, one of the latest versions of the \texttt{QUIC} specification. This comprises new requirements and requirements in \texttt{draft-18} the authors did not model. We also update the existing requirements to the new specification. We modify Ivy to support wider variables, i.e larger than 8 bytes, enabling the tool to explore more parts of the specification. 

$(2).$    We automate the use of the resulting Ivy model. We use the generated tests against seven server implementations and seven client implementations. This demonstrates that the tool can be used beyond the four original implementations tested by the authors.

$(3).$    We discuss the results we obtained and identify different types of errors in each implementation. The diversity of the results highlights contradictions and ambiguities in the 
    \texttt{QUIC} specification. 
    These findings can be used to improve the \texttt{QUIC} specification. 
    We also highlight a significant disparity in the test results and hence a difference in maturity of the protocol implementations. We provide hypothesizes of the reasons why tests failed. 


\section{The brief overview of Ivy model for \texttt{QUIC} IETF \texttt{Draft-29}}\label{sec:ivy-model}
\label{section_2}

\begin{figure*}[t]
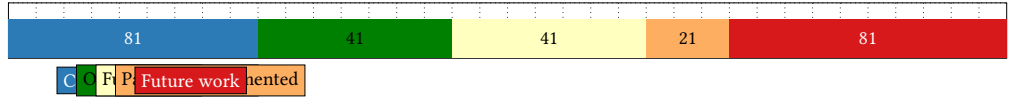

    \vspace{-1em}\par
    \centering
    \resizebox{0.75\textwidth}{!}{%
     \begin{ganttchart}[
     y unit title=0.5cm,
     y unit chart=1cm,
     vgrid,hgrid,
     title height=1,
     title label font=\bfseries\footnotesize,
     bar/.style={fill=blue},
     bar height=0.7,
     group right shift=0,
     group top shift=0.7,
     group height=.3,
     group peaks width={0.2},
     inline]{1}{36}
   
    \ganttbar[inline, bar/.append  style={fill=blue}]{\textcolor{white}{81}}{1}{9} 
    \ganttbar[inline, bar/.append  style={fill=pgreen}]{\textcolor{black}{41}}{10}{16}
    \ganttbar[inline, bar/.append  style={fill=green}]{\textcolor{black}{41}}{17}{23}
    \ganttbar[inline, bar/.append  style={fill=orange}]{\textcolor{black}{21}}{24}{26}
    \ganttbar[inline, bar/.append  style={fill=red}]{\textcolor{white}{81}}{27}{36}         
    
    \node (a) [fill=blue,draw] at ([yshift=-12pt, xshift=10pt]current bounding box.south west){\textcolor{white}{Cannot be verified}}; 
    \node (b) [fill=pgreen,draw,anchor=west] at ([xshift=10pt]a.west){\textcolor{black}{Original work}}; 
    \node (c) [fill=green,draw,anchor=west] at ([xshift=10pt]b.west){Fully implemented}; 
    \node (d) [fill=orange,draw,anchor=west] at ([xshift=10pt]c.west){\textcolor{black}{Partially implemented}}; 
    \node (e) [fill=red,draw,anchor=west] at ([xshift=10pt]d.west){\textcolor{white}{Future work}};

\end{ganttchart}
}

\caption{Tool coverage of requirements in the \texttt{QUIC} specification}
\label{must}
\end{figure*}

\begin{figure}[H]
    \centering
    \hspace*{-0.6cm}
    \includegraphics[width =1.1\columnwidth]{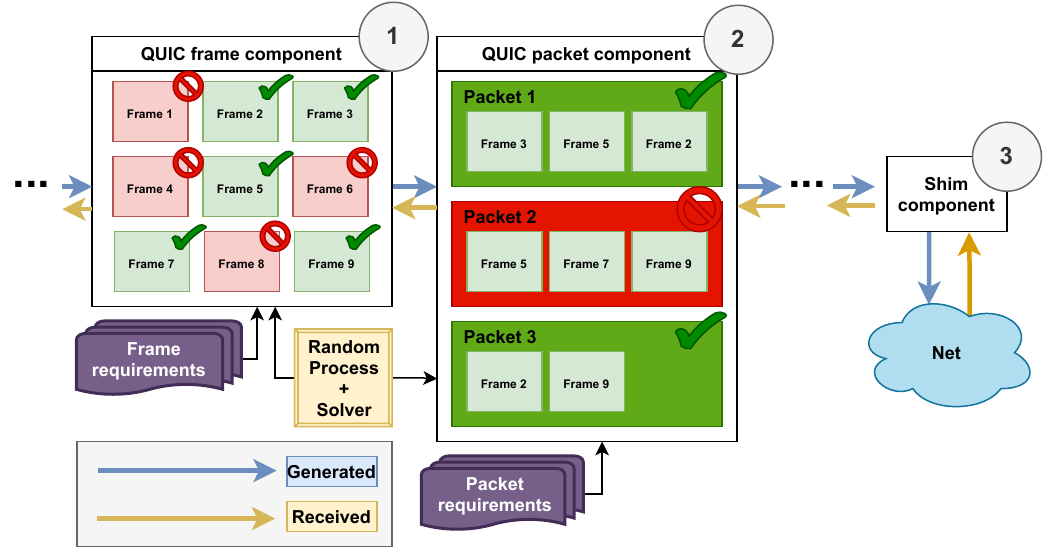}
    \caption{The \texttt{QUIC frame component} (resp. \texttt{QUIC packet component}) randomly generates frames (resp. packets) meeting its requirements, selects some of them and transmit them to the next node.
    The \texttt{shim} is the last node: it is the interface between the abstract representation of the model and the network.
    Valid \texttt{QUIC} transmissions are generated from left to right while received transmission are verified from right to left.}
    
    \label{fig:ncm}
\end{figure}

Concretely, a formal Ivy model of the protocol is defined as a set of components/nodes linked by the relationship between their input/output. Such model represents an \textit{abstraction} of the whole specification entities. Each component represents a part or a layer of the \texttt{QUIC} stack. This includes, for example, the frame layer \textcircled{1} or the packet layer \textcircled{2} presented in Figure \ref{fig:ncm}. To monitor the execution of the protocol, we define a component capturing the incoming packets, called the \texttt{shim} \textcircled{3}. 
For each received packet, the \texttt{shim} calls the \texttt{packet\_event} action, an Ivy procedure, raising an error if a requirement is violated. The \texttt{packet\_event} action contains all the requirements directly linked to the \texttt{QUIC} packet specification. It checks, for instance, that the packet number always increases. The frames are also handled similarly with their corresponding action. In Figure \ref{fig:ncm}, we can actually see that the set of requirements is directly linked to the packet component \textcircled{2}. 
The \texttt{shim} is also in charge of translating packets from Ivy to the network.


We now briefly describe our contributions to the Ivy model of \texttt{QUIC}'s specification. Our first contribution is to update this model to \texttt{draft-29} of the protocol. This is important to validate new requirements from the specification. As an example, we adapt the \texttt{shim} to the new wire format. 
Our second contribution is to extend the model with additional requirements for existing and new transport parameters. As an example, the \texttt{preferred\_address} transport parameter 
was not entirely defined in the original work, lacking a check for forbidden migration.  We also add the transport error codes management tests. For this, we create a different model which deliberately does not follow a given \texttt{QUIC} requirement in order to generate illegal packets/frames. They will be used to check whether an implementation reacts as expected, i.e. returns a specific transport error code, when receiving these packets/frames. Testing this feature is important since one can use the error code as an indication to fall back on \texttt{TCP}. 
Finally, we modify Ivy to support Connection IDs (CIDs) up to 16 bytes, increasing the domain of values that can be tested and thus the coverage of our test tool. It allows testing more implementations without modifying their source code to cope with this restriction. It allows us to relax the limitation of CIDs from exactly 8 bytes to any length up to 16 bytes.
Due to space constraints, our Ivy specification is available online\footnote{\url{https://github.com/AnonyQUIC/QUIC-Ivy}}.
 
In Figure \ref{must}, we estimate the coverage of requirements that can be handled with our Ivy model. We distinguish five categories. The first one contains requirements that cannot be verified because they only refer to internal state, lack of formal definition or cannot be modelled in Ivy due to the limitations of technique. The second category contains requirements defined in the original work that we updated to \texttt{draft-29}. The third category contains requirements for which we implemented a complete model in the tool. The fourth category contains requirements for which we implemented a partial model, i.e. requirements for which only a part can be verified. 
Finally, the last category contains requirements that we leave as future work. We will now detail how these requirements are verified.

\vspace*{-0.5cm}

\vspace*{-0.7cm}

\section{On Generating Test Cases}
\label{section_3}

\vspace*{-0.1cm}


Our objective is to test various implementations of both \texttt{QUIC} server and client. This is done by generating tests, i.e., sequences of packets/frames, from the formal Ivy model of the specification. The series of tests are generated with the \textit{Network-centric Compositional Testing} \cite{mcmillan2019compositional} methodology. This approach starts from a formal representation of the requirements given in the specification, that we described in Section~\ref{sec:ivy-model}, and generates traces, i.e. sequences of packets/frames, which can be sent to implementations.

The tests are distinguished by the type of frames they can generate and by the distributions of those frames. The tester can manually put weights on the different frame types to influence their distributions. The weights have a default value of 1. A higher (resp. lower) weight will increase (resp. reduce) the relative probability of the associated type of frame to be generated.
For the distributions, Ivy uses the Monte Carlo sampling method \cite{shapiro2003monte} that relies on those weights.

 We cannot decide to generate or not a type of packet since their utilization is restricted to some specific situations of the protocol execution. 


The input parameters of the test are also fixed. This includes, for example, the IP address used, the port or the transport parameters. This is also done manually albeit the process can be automated with a script.

The concretization of a test is a sequence of packets/frames that are sent to the implementation under test. The test stops when a requirement is not true (failing test), or when the connection closed as expected, i.e when all the data is successfully sent (passing test). 


Apart from the type of frames generated and the input parameters of the test, we can also refine the existing formal specification of \texttt{QUIC} with new requirements. By moving some of these requirements from the general model to the tests, we can have more flexibility in the tests. For example, we could fix the version of \texttt{QUIC} used directly in the based formal specification but this is not convenient when one wants to change the version in different tests. Other requirements can be used to forge illegal packets. In this case we add requirements such that the model is not conforming anymore to the \texttt{QUIC} specification. As the solver guarantees that the generated frames and packets follow these added requirements it will produce non-conform frames/packets.  







\vspace*{-0.2cm}

\subsection{Example}

We illustrate this approach on an example where we generate sequences of frames and packets. As explained above, each test starts from our formal specification of \texttt{QUIC}. We first manually fix the input parameters of the test. In this case, we manually set the IP addresses and the port used during the test. The tested implementation is a server. 

\begin{lstlisting}[style=myCustomIvyStyle]
# Tester address
parameter client_addr : ip.addr = 0x7f000001
parameter client_port : ip.port = 4987
# Tested implementation address
parameter server_addr : ip.addr = 0x7f000001
parameter server_port : ip.port = 4443        
\end{lstlisting}

Then, we choose the set of frames that will be generated and the distribution of these frames in the packets. In the following example, we decide to generate \texttt{ACK}, \texttt{STREAM}, \texttt{CRYPTO} and \texttt{PATH\_RESPONSE} frames. Then we set the relative weight for the \texttt{PATH\_RESPONSE} to five. Since this is more than one, it increases the relative probability to generate this frame. This is important since otherwise, the path validation of the migration could fail.

\begin{lstlisting}[style=myCustomIvyStyle]
# Allow generation of a frame
export frame.ack.handle
export frame.stream.handle
export frame.crypto.handle
export frame.path_response.handle
# Relative weight (all other weights = 1)
attribute frame.path_response.handle.weight = "5"
\end{lstlisting}
%



Finally, we also add specific requirements to the generated frames and packets by refining events already present in the model (i.e \texttt{packet\_event}). A requirement is indicated in Ivy with the keyword \texttt{require} and is followed by a condition. Below, we find a simple example where we randomly generate an \texttt{Initial} packet (\texttt{packet\_event}) with an \textit{invalid} "Token" field, in this case we expect a non-zero length token (see line 8). 

\begin{lstlisting}[style=myCustomIvyStyle]
# Action already present in the model
before packet_event(src:ip,dst:ip,pkt:quic_packet){
    if _generating {  # Wrong field generation
        # Applied only on generated packet
        # [...]  new features
        require ~(pkt.token.end = 0); 
    };
    # new feature (for configuration purpose)
    require pkt.long -> pkt.pversion = 0xff00001d 
}
\end{lstlisting}

As this situation is a non-conform one, we expect to receive a \texttt{PROTOCOL\_VIOLATION} error code in a \texttt{CONNECTION\_\-CLOSE} frame. This can be modelled with the special instruction \texttt{\_finalize} that allows us to check whether some requirements are eventually fulfilled.

\begin{lstlisting}[style=myCustomIvyStyle]
export action _finalize = {
    require is_invalid_token  |  ~handshake_done;
}
\end{lstlisting}


We will now give some details on tests made to stress the implementations  of the server and the client. Observe that the main reason to make a distinction between those two entities is that some parameters of the client and of the server are incompatible. As an example, the \texttt{preferred\_\-address} transport parameter is only available at the server side. Moreover, the generation of some frame such as the \texttt{HANDSHAK\-E\_DONE} is forbidden to the client.

\vspace*{-0.2cm}

\subsection{On testing server implementations}
\label{server_impl_test}

%



We have generated 23 tests to stress seven implementations of \texttt{QUIC} servers. Some of those tests are briefly described in Table \ref{tab:my-table}.



\begin{table}[H]
\centering
\resizebox{\columnwidth}{!}{%
\begin{tabular}{|ll|}
\hline
\multicolumn{2}{|l|}{\cellcolor[HTML]{333333}{\color[HTML]{FFFFFF} \textbf{quic\_server\_test\_stream}}}                                                      \\
\textbf{\begin{tabular}[c]{@{}l@{}}Generated\\ frames:\end{tabular}} & \begin{tabular}[c]{@{}l@{}}STREAM, ACK, PATH\_RESPONSE, CRYPTO\end{tabular}          \\ \hline
\textbf{\begin{tabular}[c]{@{}l@{}}Expected \\ output:\end{tabular}} &
  \begin{tabular}[c]{@{}l@{}}Simple test where we send request to index.html. Stop once all \\ expected data is received. No error is expected.\\ Note that we allow multiple connection migrations to test its \\support as well. Several streams are open to test the streams \\mechanics. We create bidirectional streams.\end{tabular} \\
\multicolumn{2}{|l|}{\cellcolor[HTML]{333333}{\color[HTML]{FFFFFF} \textbf{quic\_server\_test\_unkown}}}                                                      \\
\textbf{\begin{tabular}[c]{@{}l@{}}Generated\\ frames:\end{tabular}} & \begin{tabular}[c]{@{}l@{}}STREAM, ACK, PATH\_RESPONSE, CRYPTO, UNKNOWN\end{tabular} \\ \hline
\textbf{\begin{tabular}[c]{@{}l@{}}Expected \\ output:\end{tabular}} &
  \begin{tabular}[c]{@{}l@{}}Verifies handling of UNKNOWN frame type by the implementation.\\ Only allowed in application encryption level. We expect to receive\\ a FRAME\_ENCODING\_ERROR\end{tabular} \\ \hline
 \end{tabular}%
}
\label{tab:my-table}
\end{table} 

\newpage

 \begin{table}[H]
\centering
\resizebox{\columnwidth}{!}{%
\begin{tabular}{|ll|}
\hline
\multicolumn{2}{|l|}{\cellcolor[HTML]{333333}{\color[HTML]{FFFFFF} \textbf{quic\_server\_test\_unkown\_tp}}}                                                      \\
\textbf{\begin{tabular}[c]{@{}l@{}}Generated\\ frames:\end{tabular}} & \begin{tabular}[c]{@{}l@{}}STREAM, ACK, PATH\_RESPONSE, CRYPTO\end{tabular}          \\ \hline
\textbf{\begin{tabular}[c]{@{}l@{}}Expected \\ output:\end{tabular}} &
  \begin{tabular}[c]{@{}l@{}}Verifies handling of UNKNOWN transport parameter by the \\ implementation.We expect the implementation to ignore the\\  transport parameter and 
  continue the execution of the protocol \\normally.\end{tabular} \\ \hline
\end{tabular}%
}
\caption{Examples of Server Tests}
\label{tab:my-table}
\end{table}

\vspace*{-0.5cm}

Let us illustrate the approach on the \texttt{quic\_server\_te\-st\_stream} test. In this test, we generate \texttt{STREAM} frames until the number of requests is reached or until the maximal data defined during the handshake is exceeded.   Each time a generation or reception of a packet occurs, the whole specification will be tested. Moreover, as soon as one of the requirements is not satisfied, the test fails. 

\subsection{On testing client implementations}

We offer 14 tests to stress seven clients’ implementations. Most of those tests are like their server counterpart. Due to space limitations, we only detail in Table \ref{tab:violation-draft} the following two errors handling tests that are specific to the client's implementation. 

For the first test, we set the "Length" field of the token to zero, which is not allowed by the draft. For the second one, we verify that the \texttt{preferred\_address} transport parameter contains zero-length connection ID and the tested implementation throws a \texttt{TRANSPORT\_PARAMETER\_ERROR}. 

It is worth mentioning that we do not perform connection migration. Indeed, such migration is triggered by the client. Testing such feature would force us to change manually the tested implementation, which would not be natural. 



\begin{table}[H]
\centering
\resizebox{\columnwidth}{!}{%
\begin{tabular}{|ll|}
\hline
\multicolumn{2}{|l|}{\cellcolor[HTML]{333333}{\color[HTML]{FFFFFF} \textbf{quic\_client\_test\_new\_token\_error}}}     \\
\textbf{\begin{tabular}[c]{@{}l@{}}Generated\\ frames:\end{tabular}} & STREAM, ACK, HANDSHAKE\_DONE, CRYPTO, NEW\_TOKEN \\ \hline
\textbf{\begin{tabular}[c]{@{}l@{}}Expected \\ output:\end{tabular}} &
  \begin{tabular}[c]{@{}l@{}}We generate of NEW\_TOKEN frames with length field set to 0 \\ and expect to receive a PROTOCOL\_VIOLATION transport error.\end{tabular} \\
\multicolumn{2}{|l|}{\cellcolor[HTML]{333333}{\color[HTML]{FFFFFF} \textbf{quic\_client\_test\_prefadd\_error}}}        \\
\textbf{\begin{tabular}[c]{@{}l@{}}Generated\\ frames:\end{tabular}} & STREAM, ACK, HANDSHAKE\_DONE, CRYPTO            \\ \hline
\textbf{\begin{tabular}[c]{@{}l@{}}Expected \\ output:\end{tabular}} &
  \begin{tabular}[c]{@{}l@{}}We send the preferred\_address transport parameter with a zero length \\ CID field and we expect to receive a   \\TRANSPORT\_PARAMETER\_ERROR.\end{tabular} \\ \hline
\end{tabular}%
}
\caption{Client specific tests}
\label{tab:violation-draft}
\end{table}


\vspace*{-0.8cm}

\section{Results}




We run 100 iterations of each test on each implementation. This is to allow diverse behaviour to be captured and evaluated by the tool. Remember that each iteration generates different packets traces following the random generation approach described above. Each test was run in localhost, thus we can assume a perfect link. The results obtained against \texttt{QUIC} servers are briefly presented in Table \ref{tab:server-glob}. Those for \texttt{QUIC} clients are given in Table \ref{tab:client-glob}.  There, each column contains the results for a specific \texttt{QUIC} implementation. Each row contains the results of a test. The success rate in each cell represents the percentage of test runs satisfying all the specification requirements.

Observe that the interpretation of statistical models based on our results could be biased by the distribution of some errors.
Indeed, a few requirements could be violated for a large number of test iterations. This would lead the test to have a very low success rate while the implementation could meet all the other requirements checked by this test. For example, the \texttt{picoquic} implementation does not support migration properly. It lowers the success rate of the \texttt{stream} test to around 50\% while in practice, only one requirement is violated.

In fact, those success rates depict an insight of the implementations maturity.
This shall not be seen as a problem, as our objective is to detect errors in implementations rather than reasoning statistically.
Indeed, the detailed results express the requirements which are not met. This can give some clues for the debugging.

\begin{table}
\centering
\scriptsize
\resizebox{\columnwidth}{!}{%
\setlength\tabcolsep{2pt}
\begin{tabular}{|l|ccccccc|}
\hline
\multicolumn{1}{|c|}{} &
  \multicolumn{1}{c|}{\rotatebox{90}{\textbf{quinn} \cite{quinn}}} &
  \multicolumn{1}{c|}{\rotatebox{90}{\textbf{mvfst} \cite{mvfst}}} &
  \multicolumn{1}{c|}{\rotatebox{90}{\textbf{picoquic} \cite{picoquic}}} &
  \multicolumn{1}{c|}{\rotatebox{90}{\textbf{quic-go} \cite{quic-go}}} &
  \multicolumn{1}{c|}{\rotatebox{90}{\textbf{aioquic} \cite{aioquic}}} &
  \multicolumn{1}{c|}{\rotatebox{90}{\textbf{quant} \cite{quant}}} &
  \rotatebox{90}{\textbf{quiche}  \cite{quiche}} \\ \hline
\textbf{stream} &
  \cellcolor[HTML]{00ff55} 79\%  & 
  
\cellcolor[HTML]{FE0000} 6\%  & 
\cellcolor[HTML]{ffbb33} 56\%  &  
\cellcolor[HTML]{009933} 95\%  & 
\cellcolor[HTML]{ff6666} 18\%  & 
\cellcolor[HTML]{ff6666} 12\%  & 
  \cellcolor[HTML]{009933}97\% \\ \cline{1-1}
\textbf{max} &
  \cellcolor[HTML]{009933} 85\%  & 
  
\cellcolor[HTML]{FE0000} 3\%  & 
\cellcolor[HTML]{ffbb33} 47\%  & 
\cellcolor[HTML]{e69900} 39\%  & 
\cellcolor[HTML]{e69900} 27\%  & 
\cellcolor[HTML]{ff6666} 21\%  &
  \cellcolor[HTML]{009933}96\% \\ \cline{1-1}
\textbf{reset\_stream} &
  \cellcolor[HTML]{e69900} 29\%  & 
  
\cellcolor[HTML]{FE0000} 7\%  & 
\cellcolor[HTML]{ffff00} 61\%  & 
\cellcolor[HTML]{009933} 100\%  & 
\cellcolor[HTML]{ff6666} 24\%  & 
\cellcolor[HTML]{FE0000} 5\%  & 
  \cellcolor[HTML]{009933}98\% \\ \cline{1-1}
\textbf{connection\_close} &
  \cellcolor[HTML]{009933} 95\%  &
  
\cellcolor[HTML]{e69900} 37\%  & 
\cellcolor[HTML]{00ff55} 81\%  & 
\cellcolor[HTML]{ffff00} 63\%  & 
\cellcolor[HTML]{00ff55} 78\%  & 
\cellcolor[HTML]{e69900} 40\%  & 
  \cellcolor[HTML]{009933}100\% \\ \cline{1-1}
\textbf{stop\_sending} &
  \cellcolor[HTML]{009933}100\% &
  
\cellcolor[HTML]{FE0000} 4\%  & 
\cellcolor[HTML]{ffbb33} 48\%  & 
  \cellcolor[HTML]{F8A102}33\% &
\cellcolor[HTML]{e69900} 33\%  & 
\cellcolor[HTML]{FE0000} 8\%  & 
  \cellcolor[HTML]{009933}96\% \\ \cline{1-1}
\textbf{accept\_maxdata} &
  \cellcolor[HTML]{00ff55} 77\%  & 
  
\cellcolor[HTML]{ff6666} 12\%  & 
\cellcolor[HTML]{ffbb33} 50\%  & 
\cellcolor[HTML]{ffff00} 68\%  & 
\cellcolor[HTML]{ffbb33} 43\%  & 
\cellcolor[HTML]{ff6666} 21\%  & 
  \cellcolor[HTML]{009933}96\% \\ \cline{1-1}
  
  \textbf{unknown} &
  \cellcolor[HTML]{009933}95\% &
  \cellcolor[HTML]{009933}99\% &
  \cellcolor[HTML]{009933}99\% &
  \cellcolor[HTML]{009933}96\% &
  \cellcolor[HTML]{FE0000}{0\%} &
  \cellcolor[HTML]{FE0000}{0\%} &
  \cellcolor[HTML]{009933}100\% \\ \cline{1-1}
\textbf{unkown\_tp} &
\cellcolor[HTML]{00ff55} 84\%  & 
\cellcolor[HTML]{ffbb33} 59\%  & 
\cellcolor[HTML]{009933} 98\%  & 
\cellcolor[HTML]{009933} 100\%  & 
\cellcolor[HTML]{ffff00} 68\%  & 
\cellcolor[HTML]{009933} 100\%  & 
\cellcolor[HTML]{009933} 96\%  \\  \cline{1-1}
\textbf{double\_tp\_err} &
  \cellcolor[HTML]{FE0000}{0\%} &
  \cellcolor[HTML]{FE0000} 0\%  & 
  \cellcolor[HTML]{009933}100\% &
  \cellcolor[HTML]{009933}100\% &
  \cellcolor[HTML]{FE0000}{0\%} &
  \cellcolor[HTML]{FE0000}3\% &
  \cellcolor[HTML]{009933}100\% \\ \cline{1-1}
\textbf{tp\_err} &
  \cellcolor[HTML]{009933}100\% &
  \cellcolor[HTML]{009933}100\% &
  \cellcolor[HTML]{FE0000}0\% &
  \cellcolor[HTML]{009933}100\% &
  \cellcolor[HTML]{FE0000}0\% &
  \cellcolor[HTML]{FE0000}0\% &
  \cellcolor[HTML]{FE0000}0\% \\ \cline{1-1}
\textbf{tp\_acticoid\_err} &
  \cellcolor[HTML]{009933}100\% &
  \cellcolor[HTML]{FE0000}0\% &
  \cellcolor[HTML]{FE0000}0\% &
  \cellcolor[HTML]{FE0000}0\% &
  \cellcolor[HTML]{FE0000}0\% &
  \cellcolor[HTML]{009933}100\% &
  \cellcolor[HTML]{FE0000}0\% \\ \cline{1-1}
\textbf{no\_icid\_err} &
  \cellcolor[HTML]{009933}100\% &
  \cellcolor[HTML]{009933}100\% &
  \cellcolor[HTML]{009933}100\% &
  \cellcolor[HTML]{009933}100\% &
  \cellcolor[HTML]{FE0000}0\% &
  \cellcolor[HTML]{FE0000}0\% &
  \cellcolor[HTML]{FE0000}0\% \\ \cline{1-1}
  \textbf{token\_err} &
  \cellcolor[HTML]{009933}100\% &
  \cellcolor[HTML]{009933}98\% &
  \cellcolor[HTML]{009933}100\% &
  \cellcolor[HTML]{009933}100\% &
  \cellcolor[HTML]{009933}{\color[HTML]{333333} 100\%} &
  \cellcolor[HTML]{009933}{\color[HTML]{333333} 100\%} &
  \cellcolor[HTML]{009933}99\% \\ \cline{1-1}
\textbf{new\_token\_err} & 
\cellcolor[HTML]{009933} 100\%  & 
\cellcolor[HTML]{FE0000} 0\%  & 
\cellcolor[HTML]{FE0000} 0\%  & 
\cellcolor[HTML]{00ff55} 84\%  & 
\cellcolor[HTML]{009933} 100\%  & 
\cellcolor[HTML]{FE0000} 0\%  & 
\cellcolor[HTML]{FE0000} 0\%  \\ \cline{1-1}
\textbf{handshake\_done\_err} &
  \cellcolor[HTML]{009933}100\% &
  \cellcolor[HTML]{009933}92\% &
  \cellcolor[HTML]{009933}89\% &
  \cellcolor[HTML]{FE0000}0\% &
  \cellcolor[HTML]{009933}86\% &
  \cellcolor[HTML]{FE0000}2\% &
  \cellcolor[HTML]{00ff55}77\% \\ \cline{1-1}
 
\textbf{newcid\_err} & 
\cellcolor[HTML]{00ff55} 81\% & 
\cellcolor[HTML]{009933} 85\%  & 
\cellcolor[HTML]{009933} 100\%  & 
\cellcolor[HTML]{FE0000} 9\%  & 
\cellcolor[HTML]{ffff00} 68\%  & 
\cellcolor[HTML]{009933} 93\%  & 
\cellcolor[HTML]{009933} 91\% \\ \cline{1-1}
\textbf{max\_limit\_err} & 
\cellcolor[HTML]{ffbb33} 49\%  & 
\cellcolor[HTML]{ffbb33} 41\%  & 
\cellcolor[HTML]{009933} 100\%  & 
\cellcolor[HTML]{FE0000} 0\%  & 
\cellcolor[HTML]{ffbb33} 41\%  & 
\cellcolor[HTML]{ff6666} 16\%  & 
\cellcolor[HTML]{FE0000} 0\%   \\ \cline{1-1}

\textbf{blocked\_err} &
  \multicolumn{1}{c}{\cellcolor[HTML]{34FF34}70\%} &
  \multicolumn{1}{c}{\cellcolor[HTML]{FE0000}0\%} &
  \multicolumn{1}{c}{\cellcolor[HTML]{FE0000}0\%} &
  \multicolumn{1}{c}{\cellcolor[HTML]{34FF34}75\%} &
  \multicolumn{1}{c}{\cellcolor[HTML]{FE0000}{0\%}} &
  \multicolumn{1}{c}{\cellcolor[HTML]{FE0000}{0\%}} &
  \multicolumn{1}{c|}{\cellcolor[HTML]{009933}100\%} \\ \cline{1-1}
\textbf{retirecid\_err} & 
\cellcolor[HTML]{009933} 87\%  & 
\cellcolor[HTML]{FE0000} 0\%  & 
\cellcolor[HTML]{009933} 86\%  & 
\cellcolor[HTML]{009933} 85\%  & 
\cellcolor[HTML]{FE0000} 0\%  & 
\cellcolor[HTML]{FE0000} 0\%  & 
\cellcolor[HTML]{FE0000} 0\%  \\ \cline{1-1}
\textbf{stream\_limit\_err} &
\cellcolor[HTML]{009933} 100\% &
\cellcolor[HTML]{ffbb33} 63\% &
\cellcolor[HTML]{009933} 99\%  & 
\cellcolor[HTML]{009933} 98\%  & 
\cellcolor[HTML]{009933} 99\%  & 
\cellcolor[HTML]{FE0000} 10\%  & 
\cellcolor[HTML]{FE0000} 0\%  \\ \cline{1-1}
\textbf{newcid\_length\_err} &
\cellcolor[HTML]{00ff55} 84\%  & 
\cellcolor[HTML]{FE0000} 0\%  & 
\cellcolor[HTML]{FE0000} 2\%  & 
\cellcolor[HTML]{00ff55} 81\%  & 
\cellcolor[HTML]{FE0000} 0\%  & 
\cellcolor[HTML]{FE0000} 0\%  & 
\cellcolor[HTML]{009933} 91\%  \\ \cline{1-1}
\textbf{newcid\_rtp\_err} & 
\cellcolor[HTML]{009933} 91\%  & 
\cellcolor[HTML]{FE0000} 0\%  & 
\cellcolor[HTML]{FE0000} 0\%  & 
\cellcolor[HTML]{009933} 90\%  & 
\cellcolor[HTML]{FE0000} 0\%  & 
\cellcolor[HTML]{FE0000} 0\%  & 
\cellcolor[HTML]{FE0000} 0\%  \\ \cline{1-1}
\textbf{max\_err} & 
\cellcolor[HTML]{FE0000} 0\%  & 
\cellcolor[HTML]{009933} 90\% &
\cellcolor[HTML]{009933} 100\%  & 
\cellcolor[HTML]{FE0000} 0\%  & 
\cellcolor[HTML]{FE0000} 0\%  & 
\cellcolor[HTML]{FE0000} 0\%  & 
\cellcolor[HTML]{FE0000} 0\%  \\ \hline
\end{tabular}%
}
\caption{Server - Successful test ratio}
\label{tab:server-glob}
\vspace*{-0.9cm}
\end{table}






\begin{table} 
\centering
\scriptsize
\resizebox{\columnwidth}{!}{%
\setlength\tabcolsep{2pt}
\begin{tabular}{|l|ccccccc|}
\hline
\multicolumn{1}{|c|}{} &
  \multicolumn{1}{c|}{\rotatebox{90}{\textbf{quinn} \cite{quinn}}} &
  \multicolumn{1}{c|}{\rotatebox{90}{\textbf{picoquic} \cite{picoquic}}} &
  \multicolumn{1}{c|}{\rotatebox{90}{\textbf{quic-go} \cite{quic-go}}} &
  \multicolumn{1}{c|}{\rotatebox{90}{\textbf{aioquic} \cite{aioquic}}} &
  \multicolumn{1}{c|}{\rotatebox{90}{\textbf{quant} \cite{quant}}} &
  \multicolumn{1}{c|}{\rotatebox{90}{\textbf{quiche} \cite{quiche}}} &
  \rotatebox{90}{\textbf{lsquic}  \cite{lsquic}} \\ \hline
\textbf{stream} &
  \cellcolor[HTML]{009933} 99\% &
  \cellcolor[HTML]{ffbb33} 51\% &
  \cellcolor[HTML]{009933} 100\% &
  \cellcolor[HTML]{009933} 97\% &
  \cellcolor[HTML]{00ff55} 85\% &
  \cellcolor[HTML]{ffbb33} 52\% &
  \cellcolor[HTML]{009933} 92\% \\ \cline{1-1}
\textbf{max} &
  \cellcolor[HTML]{009933} 100\% &
  \cellcolor[HTML]{ff6666} 15\% &
  \cellcolor[HTML]{009933} 100\% &
  \cellcolor[HTML]{009933} 98\% &
  \cellcolor[HTML]{00ff55} 85\% &
  \cellcolor[HTML]{e69900} 34\% &
  \cellcolor[HTML]{009933} 100\% \\ \cline{1-1}
\textbf{accept\_maxdata} &
  \cellcolor[HTML]{009933} 100\% &
  \cellcolor[HTML]{009933} 93\% &
  \cellcolor[HTML]{009933} 100\% &
  \cellcolor[HTML]{009933} 97\% &
  \cellcolor[HTML]{009933} 95\% &
  \cellcolor[HTML]{00ff55} 82\% &
  \cellcolor[HTML]{00ff55} 83\% \\ \cline{1-1}
\textbf{unkown} &
  \cellcolor[HTML]{009933} 100\% &
  \cellcolor[HTML]{009933} 96\% &
  \cellcolor[HTML]{009933} 99\% &
  \cellcolor[HTML]{FE0000} 0\% &
  \cellcolor[HTML]{FE0000} 0\% &
  \cellcolor[HTML]{009933} 100\% &
  \cellcolor[HTML]{FE0000} 0\% \\ \cline{1-1}
\textbf{tp\_unkown} &
  \cellcolor[HTML]{009933} 100\% &
  \cellcolor[HTML]{e69900} 34\% &
  \cellcolor[HTML]{009933} 99\% &
  \cellcolor[HTML]{009933} 99\% &
  \cellcolor[HTML]{009933} 100\% &
  \cellcolor[HTML]{009933} 99\% &
  \cellcolor[HTML]{009933} 96\% \\ \cline{1-1}
\textbf{double\_tp\_error} &
  \cellcolor[HTML]{FE0000} 0\% &
  \cellcolor[HTML]{009933} 100\% &
  \cellcolor[HTML]{009933} 100\% &
  \cellcolor[HTML]{FE0000} 0\% &
  \cellcolor[HTML]{FE0000} 0\% &
  \cellcolor[HTML]{FE0000} 0\% &
  \cellcolor[HTML]{FE0000} 0\% \\ \cline{1-1}
\textbf{tp\_error} &
\cellcolor[HTML]{FE0000} 0\%  & 
\cellcolor[HTML]{FE0000} 0\%  & 
\cellcolor[HTML]{009933} 100\%  & 
\cellcolor[HTML]{FE0000} 0\%  & 
\cellcolor[HTML]{FE0000} 0\%  & 
\cellcolor[HTML]{FE0000} 0\%  & 
\cellcolor[HTML]{FE0000} 0\% \\ \cline{1-1}
\textbf{tp\_acticoid\_error} &
  \cellcolor[HTML]{FE0000} 0\% &
  \cellcolor[HTML]{FE0000} 0\% &
  \cellcolor[HTML]{FE0000} 0\% &
  \cellcolor[HTML]{FE0000} 0\% &
  \cellcolor[HTML]{009933} 100\% &
  \cellcolor[HTML]{FE0000} 0\% &
  \cellcolor[HTML]{FE0000} 0\% \\ \cline{1-1}
\textbf{no\_ocid} &
  \cellcolor[HTML]{FE0000} 0\% &
  \cellcolor[HTML]{009933} 100\% &
  \cellcolor[HTML]{009933} 100\% &
  \cellcolor[HTML]{FE0000} 0\% &
  \cellcolor[HTML]{FE0000} 0\% &
  \cellcolor[HTML]{FE0000} 0\% &
  \cellcolor[HTML]{FE0000} 0\% \\ \cline{1-1}
\textbf{tp\_prefadd\_error} &
  \cellcolor[HTML]{FE0000} 0\% &
  \cellcolor[HTML]{009933} 100\% &
  \cellcolor[HTML]{FE0000} 0\% &
  \cellcolor[HTML]{FE0000} 0\% &
  \cellcolor[HTML]{FE0000} 0\% &
  \cellcolor[HTML]{FE0000} 0\% &
  \cellcolor[HTML]{FE0000} 0\% \\ \cline{1-1}
\textbf{blocked\_error} &
  \cellcolor[HTML]{009933} 99\% &
  \cellcolor[HTML]{FE0000} 0\% &
  \cellcolor[HTML]{009933} 97\% &
  \cellcolor[HTML]{FE0000} 0\% &
  \cellcolor[HTML]{FE0000} 0\% &
  \cellcolor[HTML]{009933} 91\% &
  \cellcolor[HTML]{009933} 98\% \\ \cline{1-1}
\textbf{retirecoid\_error} &
  \cellcolor[HTML]{009933} 99\% &
  \cellcolor[HTML]{009933} 99\% &
  \cellcolor[HTML]{009933} 100\% &
  \cellcolor[HTML]{FE0000} 0\% &
  \cellcolor[HTML]{FE0000} 0\% &
  \cellcolor[HTML]{FE0000} 0\% &
  \cellcolor[HTML]{009933} 98\% \\ \cline{1-1}
\textbf{new\_token\_error} &
  \cellcolor[HTML]{009933} 98\% &
  \cellcolor[HTML]{009933} 94\% &
  \cellcolor[HTML]{009933} 96\% &
  \cellcolor[HTML]{FE0000} 1\% &
  \cellcolor[HTML]{FE0000} 0\% &
  \cellcolor[HTML]{00ff55} 87\% &
  \cellcolor[HTML]{009933} 100\% \\ \cline{1-1}
\textbf{limit\_max\_error} &
  \cellcolor[HTML]{FE0000} 0\% &
  \cellcolor[HTML]{00ff55} 88\% &
  \cellcolor[HTML]{FE0000} 0\% &
  \cellcolor[HTML]{FE0000} 0\% &
  \cellcolor[HTML]{00ff55} 81\% &
  \cellcolor[HTML]{FE0000} 0\% &
  \cellcolor[HTML]{FE0000} 0\% \\ \hline
\end{tabular}%
}
\caption{Client - Successful test ratio}
\label{tab:client-glob}
\vspace*{-1cm}
\end{table}



We will now present the two main categories of problems found. The first one concerns all problems related to migration and the second one concerns those related to the transport errors. Note that due to the page limitation, we did not include all the results but these will be available in a technical version of the paper.

\subsection{Migration issues}


For the tests with only legal frames/packets, we do not detect many different errors. Most of the problems are linked to acknowledgement of acknowledgements and the path validation during a migration. 

The first problem never leads to a crash and no major security issues were detected. We supposed that it is also linked to migration since this error amplifies when migration is allowed. This is enhanced when we compare the result of the client tests and the server tests. This error is less present when the client is tested. As a reminder, the server tests allow multiple connection migrations while and client tests disable it.

However the migration problems are more serious. Many security considerations are involved in guaranteeing authentication, confidentiality and integrity of the messages exchanged between endpoints with the migration in general. Many attacks are linked to the migration such as
the "Peer Address Spoofing", "On-Path Address Spoofing" or the "Off-Path Packet
Forwarding" describe in the \texttt{QUIC} specification. It is important for the path and
address validation to be well implemented.

\vspace*{-0.1cm}
\subsubsection{Migration issues: Case study I}
\hfill\\

\vspace*{-0.3cm}
An example of problem with the migration was highlighted with \texttt{mvfst} implementation. The migration is considered invalid when we start it just after receiving the \texttt{HANDSHAKE\_DONE} and the \texttt{disable\_active\_migration} transport parameter is not set according to Wireshark and our tool. In this case we acknowledge the frame in the migrated connection. According to Section 9 of the specifications, an endpoint cannot migrate before the handshake is confirmed. From the client point of view, we consider the handshake as confirmed when we received the \texttt{HANDSHAKE\_DONE} frame.

For the server, the handshake is confirmed when the handshake is complete. It happens when the \texttt{TLS} stacks had both sent a "Finished" message and verified the peer’s "Finished" message. 

Thus, if a client migrates just after it receives a \texttt{HANDSHAKE\-\_DONE}, then the client respects the draft. The server has to accept the migration before it receives the \texttt{ACK} for the \texttt{HANDSHAKE\_DONE}. 

However the \texttt{mvfst} implementation server only knows that the client considers the handshake as confirmed when it received the \texttt{ACK} for \texttt{HANDSHAKE\_DONE}. We suppose that \texttt{mvfst} wants to be sure that the client has received the \texttt{HANDSHAKE\_DONE} before allowing it to migrate.  Moreover, in the case where the frame is acknowledged in the original connection, this problem does not arise. This is consistent with our hypothesis. 



\subsubsection{Migration issues: Case study II}
\hfill\\

Another problem reflected in our model is that polysemous requirements in the draft lead to different valid formal interpretations. This means that they are ambiguous. The best example is with the
highest-numbered non-probing packet to which an endpoint should send its packets. Consider the following statement:

\vspace*{-0.1cm}

\epigraph{ "\textit{An endpoint only changes the address that it sends packets to in\\
response to the highest-numbered non-probing packet.  This ensu-\\
res that an endpoint does not send packets to an old peer address\\in the case that it receives reordered packets}."}{Draft 29 of \texttt{QUIC} section 9.3.}
\vspace*{-0.1cm}

According to this rule, we should send the packet to the highest-numbered non-probing packet. A non-probing packet is a packet containing any other frame than the probing frames which are \texttt{PATH\_CHALLENGE}, \texttt{PATH\_RESPONSE}, \texttt{NEW\_CO}- \texttt{NNECTION\_ID}, and \texttt{PADDING} frames.  As we did not have an indication about the encryption level to consider, we used the highest-numbered non-probing packet among all possible ones. Since migration is allowed only after the handshake completed, we should probably consider only the encryption level for the 0-RTT Protected or short header packets. It is not written explicitly.

However, in the same paragraph, we can find the following indication:

\vspace*{-0.1cm}

\epigraph{ "\textit{Receiving a packet from a new peer address containing a non-probing frame
indicates that the peer has migrated to that address. In response to such a packet, an
endpoint MUST start sending subsequent packets to the new peer address and MUST
initiate path validation (Section 8.2) to verify the peer’s ownership of the
unvalidated address}."}{Draft 29 of \texttt{QUIC} section 9.3.}

\vspace*{-0.1cm}


\noindent
If we use this rule, then we must send responses of non-probing frames to the new peer address. When we consider the highest-numbered non-probing packet among all the possible encryption level, most of the implementation do not pass the requirement. This is the case with \texttt{picoquic}. However when we consider the highest-numbered non-probing packet among the application encryption level, some of the implementations that do not pass with the above interpretation manage to pass the requirement. This is again the case for \texttt{picoquic}. We show that both chosen interpretations of the requirement, which are valid from a formal point of view, lead to different results. Even if the best solution seems trivial for a network specialist, the ambiguity should be removed.


\vspace*{-0.1cm}
\subsection{Transport error code issues}

\vspace{-0.1cm}

Concerning the management of the transport error code, we can see that some implementations respect the specification better than others. Various types of errors have been observed. We faced some implementations that do not implement a requirement in the specification. For example, the transport error codes management is almost completely implemented in some implementations (e.g \texttt{quinn}) and barely implemented in other ones (e.g. \texttt{quant}) as can see in Table \ref{tab:server-glob}. As the application layer can use those error codes to decide whether to fall back on a \texttt{TCP} connection, the transport error codes management is important. 


Another type of problem is when implementations use the wrong encryption level. This is mostly the case for \texttt{aioquic} or \texttt{quinn}. Those two implementations use the \texttt{1-RTT} encryption scheme to return the error. However this encryption level is not allowed at this time. 

Some implementations do not check if a specific frame is allowed by their peers. For example, the \texttt{NEW\_TOKEN} frame (sent by \texttt{picoquic} and \texttt{mvfst}) is not allowed to be sent by the client. These implementations do not report any \texttt{PROTOCOL\_VIOLATION} that does not conform to the specification, neither local error. 

We also observed that some implementation send error codes with the wrong error message. As an example, when we send \texttt{HANDSHAKE\_DONE} frame to the \texttt{quic-go} server, which is not allowed, it closes the connection with the message error reporting that this frame is not allowed because of the encryption level. This is not correct in this case.


Some implementations return the wrong error code. An example is \texttt{mvfst} when we send \texttt{NEW\_CONNECTION\_ID} frame with an invalid field, \texttt{mvfst} detects the correct problem but return the wrong error code. We know it detects the problem since the error message describe the fact that the field is invalid.


Some implementations like \texttt{quant} detect the error locally without reporting an error. This is not conforming to the specification. However, when we analyse the latest version, that problem has been solved as presented in Table \ref{tab:quant}.


\vspace*{-0.3cm}


\begin{table}[H]
\centering
\small
\begin{tabular}{|l|cc|}
\hline
\multicolumn{1}{|c|}{\cellcolor[HTML]{FFFFFF}{\color[HTML]{FFFFFF}}} &
  \multicolumn{1}{c|}{\textbf{quant \textit{29}}} &
 \textbf{quant \textit{master}} \\ \hline
\textbf{double\_tp\_error} &
  \cellcolor[HTML]{FE0000}3\% &
  \cellcolor[HTML]{009933}100\% \\ \cline{1-1}
\textbf{tp\_error} &
  \cellcolor[HTML]{FE0000}0\% &
\cellcolor[HTML]{009933} 100\%   \\ \cline{1-1}
\textbf{tp\_acticoid\_error} &
  \cellcolor[HTML]{009933}100\% &
\cellcolor[HTML]{009933} 100\%  \\ \cline{1-1}
\textbf{no\_icid\_error } &
  \cellcolor[HTML]{FE0000}0\% &
\cellcolor[HTML]{009933} 100\% \\ \hline
\end{tabular}%
\caption{Quant transport parameter: before/after}
\vspace*{-0.5cm}
\label{tab:quant}
\end{table}








\section*{Conclusion and Future Work}


In this paper, we applied the Ivy modelization framework to \texttt{QUIC} and produced a model verifying a large part of the requirements of the \texttt{QUIC} specification. The model contains new and updated requirements from the original work. We automated the use of the model in a series of tests against \texttt{QUIC} clients and servers. We showed the diversity of results we obtained with our tool and the variety of errors encountered. In addition, the results we obtained illustrate the ability of the tool to highlight ambiguities in the specification and generate executions following different interpretations of these ambiguities. This shall be exploited to improve the IETF specification of \texttt{QUIC}.



We see several directions for future work. One of them is to include in our approach extensions and additional features of \texttt{QUIC} such as \texttt{Multipath QUIC}~\cite{10.1145/3143361.3143370}, \texttt{Forward Erasure Correction} (FEC)~\cite{michel2019quic}, the Unreliable Datagram extension \cite{pauly2018unreliable} and packets such as \texttt{Retry} and \texttt{Version Negotiation}.


\vspace*{-0.2cm}

\section*{Software artefacts}

We will release our modifications to Ivy, the updated test suite and the Docker images automating the tests~\cite{TowardsverificationofQUICextensions}\footnote{\url{https://github.com/AnonyQUIC/QUIC-Ivy}}. Here are the commits of the tested implementations:

\vspace*{-0.3cm}

\begin{table}[H]
\centering
\small

\resizebox{\columnwidth}{!}{%
\begin{tabular}{|c|c|}
\hline
\textbf{picoquic (C)}  \cite{picoquic} & \textbf{lsquic}         \cite{lsquic}           \\
ad23e6c3593bd987dcd8d74fc9f528f2676fedf4   & v2.29.4                                      \\ \cline{2-2} 
\textbf{picotls}  \cite{picotls}               & \textbf{boringssl}           \cite{boringssl}      \\
47327f8d032f6bc2093a15c32e666ab6384ecca2 & a2278d4d2cabe73f6663e3299ea7808edfa306b9 \\ \hline
\textbf{quic-go (golang)}    \cite{quic-go}    & \textbf{quinn (rust)}     \cite{quinn}         \\
v0.20.0                                    & 0.7.0                                        \\ \hline
\textbf{aioquic (python)}   \cite{aioquic}     & \textbf{quiche (rust)}     \cite{quiche}        \\
0.9.3                                      & 0.7.0                                        \\ \hline
\textbf{quant (C)}    \cite{quant}           & \textbf{mvfst (facebook/C++)}     \cite{mvfst} \\
29                                         & 36111c1                                      \\ \hline
\end{tabular}%
}
\caption{Draft 29 tested implementations}
\label{tab:QUIC-Implementations}
\end{table}




\newpage

\bibliographystyle{ACM-Reference-Format}
\bibliography{reference}

\newpage

\end{document}